# Two-electron quantum dot in a magnetic field: Analytic solution for finite potential model

S. Chaudhuri

Department of Chemistry and Physics

Monmouth University

400 Cedar Avenue

West Long Branch, NJ 07764

schaudhu@monmouth.edu

The energy levels of two interacting electrons in a 2D quantum dot confined by a finite Gaussian potential and subjected to a uniform magnetic field perpendicular to the plane of the dot are studied. Analytic results are obtained for the energy spectrum of the two-electron system as well as for a single electron. The magnetic field at which the ground state transitions from the spin-singlet to the spin-triplet state for the two-electron system is calculated and compared with the experimental data and previous theoretical results for the infinite harmonic confining potential. The ground state energy of the two-electron system as a function of the magnetic field is calculated and compared with the experimental data.

## I. INTRODUCTION

Soon after the development of the quantum theory two-dimensional systems in a magnetic field have been of interest to physicists. Fock [1] and subsequently Darwin [2] derived the energy levels of a single particle confined in a two-dimensional harmonic potential and subjected to a uniform magnetic field. Studies of such systems intensified with the fabrication of quantum systems confined in two and all three spatial dimensions known as artificial atoms, superatoms, or quantum dots, more than thirty years ago. See Reference [3] for an extensive list of references on the topic.

Two-dimensional disc shaped quantum dots (2D QD) where the number of confined electrons and the confining potential can be controlled at will have been studied experimentally and theoretically [4-29]. Various techniques including Hartree-Fock, variational approach, numerical diagonalization of the Hamiltonian matrix, etc. have been used to calculate the energy levels of two-electron systems in a 2D QD. Most of the theoretical treatments have approximated the confining potential to be an infinite harmonic potential. The major advantage of using the harmonic potential is that the Hamiltonian can be separated in the center of mas (CM) and relative coordinates (RC). Analytic solution to the CM part of the Schrodinger equation (SE) exists. However, the RC SE including the Coulomb interaction between the two electrons is difficult and hence the various approximate techniques for analytic treatments and exact numerical techniques have been used. Bruce and Maksym [30] treated the problem with a realistic confining potential including the effect of screening due to the gate electrodes using numerical diagonalization of the many-body Hamiltonian. Adamowski et al. [29] used a Gaussian potential and its parabolic approximation to calculate the energy levels of a QD without a magnetic field using the variational approach.

Wagner et al. [8] predicted a transition for the ground state energy from the spin-singlet to the spin-triplet state as the magnetic field increases. Ashoori et al. [31] experimentally observed such transition.

However, there is a discrepancy in the experimental value of the magnetic field with the value calculated by Wagner et al. Ashoori et al. conjectured that the discrepancy may be due to the assumption of strictly parabolic potential used in the calculation. In this paper, we model the system with a Gaussian potential that should be a more realistic representation of the 2D QD and present analytic results for this potential. However, for our analytic treatment we have made a key assumption that the CM motion is confined mostly near the center of the QD. With this assumption we have obtained an analytic solution for the system. The transition magnetic field value obtained for the Gaussian potential is closer to the experimental value indicating that indeed the discrepancy may be at least partly due to the infinite harmonic potential assumption by Wagner et al.

Dineykhan and Efimov [32] developed the *oscillator representation method* (ORM) arising from ideas and methods of the quantum field theory. Using the ORM the binding energies of a number of systems with various types of potentials including the Coulomb and power-law potentials, exponentially screened Coulomb potential, logarithmic potential [32, 33], van der Waals potential [34], cavity model [35], and a two-electron quantum dot in a magnetic field [36] have been calculated. The ORM results agree very well with the results obtained by variational numerical methods and analytic methods for these potentials. In this paper, we use the ORM technique to obtain an analytic solution for the 2D QD two-electron systems confined by a Gaussian potential in a magnetic field.

The paper is organized as follows. In Section II we present the Gaussian potential we have used for our model and the approximations involved for the model. In Section II we present the results obtained by the ORM solution for the RC SE. In Section III we present the single electron ORM results. By fitting the single electron experimental data [31] with our results we obtain the parameters for the Gaussian potential. In Section IV we calculate the energy levels and the magnetic field value for the singlet-to-triplet transition with the potential parameters obtained in Section III. We compare the results with both the experimental data [31] and the results by Wagner et al. [8].

## II. GAUSSIAN CONFINING POTENTIAL MODEL

The Hamiltonian of a two-electron system in a 2D quantum dot (QD) is:

$$H = \sum_{i=1}^{2}[\frac{1}{2m^*}(\boldsymbol{p}_i + e\boldsymbol{A}(\boldsymbol{r}_i))^2 + V(\boldsymbol{r}_i)] + \frac{e^2}{4\pi\epsilon\epsilon_0}\frac{1}{|\boldsymbol{r}_1 - \boldsymbol{r}_2|} + H_s, \tag{1}$$

where $m^*$ is the effective mass, $\boldsymbol{A}$ is the vector potential corresponding to a magnetic field $\boldsymbol{B}$ perpendicular to the plane of the 2D QD, $\epsilon$ is the dielectric constant of the QD medium, and $H_s = g^*\mu_B(\boldsymbol{s}_1 + \boldsymbol{s}_2).\boldsymbol{B}$. We choose the conventional gauge for the magnetic field $\mathbf{B}$ in the z-direction described by $A = \frac{1}{2}(\boldsymbol{B} \times \boldsymbol{r})$.

We choose the confining potential to be

$$V(\boldsymbol{r}_i) = V_0\left(1 - e^{-\alpha^2 r_i^2}\right). \tag{2}$$

For $\alpha r_i \ll 1, V(\boldsymbol{r}_i) \cong V_0\alpha^2 r_i^2$. To facilitate comparison of the model with harmonic potential model we set

$$V_0\alpha^2 = \frac{1}{2}m^*\omega_0^2. \tag{3}$$

It is necessary to separate the Hamiltonian into two components, one corresponding to the (CM) and the other to the relative motion of the two-electron system in order to solve the SE analytically. That is the primary motivation to approximate the confining potential to be harmonic potential for which the Hamiltonian is separable in the two coordinates. The Hamiltonian is not separable for the Gaussian potential. With some assumptions we can, however, separate the Hamiltonian with the Gaussian potential that should provide an approximate model for realistic QDs.

The two-electron confinement potential in the CM and the relative coordinates, $\boldsymbol{R} = \frac{\boldsymbol{r_1+r_2}}{2}$, and $\boldsymbol{r} = \boldsymbol{r_1} - \boldsymbol{r_2}$, respectively, is

$$V = 2V_0\left[1 - e^{-\alpha^2(R^2+\frac{r^2}{4})}\cosh(\alpha^2 Rr\cos\theta)\right], \tag{4}$$

where $\boldsymbol{R}.\boldsymbol{r} = Rr\cos\theta$. Assuming $\alpha^2 R^2 \ll 1$, we neglect terms of order higher than $\alpha^2 R^2$ including $\alpha^4 R^2 r^2$ to obtain the confining potential as

$$V = 2V_0\left[1 - e^{-\alpha^2 r^2/4} + \alpha^2 R^2\right]. \tag{5}$$

Later in the paper we will include the $\alpha^4 R^2 r^2$ term as an average with the expectation value $\alpha^4 R^2 < n,m|r^2|n,m >$ after the wave functions $|n,m>$ are determined by solving the RC SE.

By fitting the calculated eigen energy of a single electron in a confining potential $V_0(1 - e^{-\alpha^2 r^2})$ and subjected to a perpendicular uniform magnetic field with the single electron experimental data [31]. we find the expectation value $< \alpha^2 R^2 >$ to be 0.1. This indicates that the assumption made to obtain the confining potential may be a reasonable one for realistic QDs to study the effect of finite potential barrier on the energy spectrum of two-electron systems.

With the potential in Equation (5) the Hamiltonian is separated in the CM and the relative coordinates and written as

$$H = \frac{1}{2}H_Q + 2H_q + H_s, \tag{6}$$

where $H_Q$ and $H_q$ are the Hamiltonian parts corresponding to the CM and relative coordinate motions, respectively, and written as (following the notations of Dineykhan and Nazmitdinov [36])

$$H_Q = \frac{1}{2}\left[\boldsymbol{P_Q} + \boldsymbol{A_Q}\right]^2 + \frac{1}{2}\hbar^2\omega_Q^2, \text{ and} \tag{7}$$

$$H_q = \frac{1}{2}\left[\boldsymbol{p_q} + \boldsymbol{A_q}\right]^2 + \frac{1}{2}\hbar^2\omega_q^2 + \frac{m^*\omega_0^2}{2\alpha^2}\left(1 - e^{-\frac{\alpha^2\hbar^2}{4m^*}q^2}\right) + \frac{k\sqrt{\hbar\omega_0}}{2q}, \tag{8}$$

where $\omega_Q = 2\omega_0$, $\omega_q = \omega_0/2$, $\boldsymbol{A_Q} = \boldsymbol{A(q_1)} + \boldsymbol{A(q_2)}$, $\boldsymbol{A_q} = \frac{1}{2}[\boldsymbol{A(q_1)} - \boldsymbol{A(q_2)}]$, and $\boldsymbol{A(q)} = \left(\frac{\hbar}{m^*}\right)[\boldsymbol{B}\times\boldsymbol{q}]$, $\boldsymbol{Q} = \left(\sqrt{m^*/\hbar}\right)\boldsymbol{R}$, and $\boldsymbol{q} = \left(\sqrt{m^*/\hbar}\right)\boldsymbol{r}$. The coulomb interaction constant $k = l_0/a^*$, where $l_0 = \sqrt{\hbar/m^*\omega_0}$ is a characteristic length of the harmonic oscillator of frequency $\omega_0$ and $a^* = \epsilon(m_e/m^*)a_B$ is the effective Bohr radius ($a_B = 4\pi\epsilon_0 \frac{\hbar^2}{m_e e^2}$).

The eigenfunction of the Hamiltonian, Eq. (6), then can be written in the separable coordinates **R** and **r**, as

$$\Psi = \Phi(\boldsymbol{Q})\psi(\boldsymbol{q})\chi(\boldsymbol{s_1}, \boldsymbol{s_2}). \tag{9}$$

The corresponding SEs are:

$$H_Q\Phi(\boldsymbol{Q}) = \varepsilon_Q\Phi(\boldsymbol{Q}), and \tag{10a}$$

$$H_q\psi(\boldsymbol{q}) = \varepsilon_q\psi(\boldsymbol{q}). \tag{10b}$$

The solution to the SE for the CM motion is well known. Due to the circular symmetry of the Hamiltonian, the CM as well as the relative angular momentums are conserved resulting in the azimuthal quantum numbers $M$ and $m$, respectively. Representing the radial quantum numbers for the $Q$ and $q$ by $N$ and $n$, respectively, the eigenvalues $\varepsilon_{N,M}$ are given by [1, 2]

$$\varepsilon_{N,M} = 2\hbar\omega_0\left[\gamma(2N + |M| + 1) + \frac{1}{2}Mt\right]. \tag{11}$$

where $\gamma = (1 + t^2/4)^{1/2}$, $t = \omega_c/\omega_0$, $\omega_c = eB/m^*$, $N = 0, 1, 2, \ldots$, and $M = 0, \pm 1, \pm 2, \ldots$, and the eigenfunctions are given by

$$\Phi(R) = \frac{1}{l_0}\sqrt{\frac{2}{\pi}}\left[\frac{\Gamma(N+1)}{\Gamma(N+|M|+1)}\right]^{\frac{1}{2}}e^{iM\phi_Q}e^{-x^2}x^{|M|}L_N^{|M|}(x^2), \tag{12}$$

where, $x = \frac{\sqrt{2}R}{l_0\sqrt{\gamma}}$, $L_N^{|M|}$ is the Associated Laguerre function, and $\phi_Q$ is the azimuthal angle corresponding to the CM coordinate **R.**

The total energy of the two-electron system is

$$E = E_{N,M} + E_{n,m} + E_s, \tag{13}$$

where, $E_{N,M} = \frac{1}{2}\varepsilon_{N,M}$ and $E_{n,m} = 2\varepsilon_{n,m}$.

The requirement by the Pauli exclusion principle that the total wavefunction must be antisymmetric with respect to particle permutation (i.e., $\boldsymbol{q} \to -\boldsymbol{q}$) leads to the requirement that the spin wavefunction $\chi$ must be singlet (S=0) and triplet (S=1) for even and odd $m$, respectively. The spin energy $E_{S,m}$ for the singlet and triplet states alternating with $m$ is given by

$$E_{S,m} = \hbar\omega_0 \cdot \frac{1}{4}[1 - (-1)^m]g^*\frac{m^*}{m_e}t, \tag{14}$$

where, $g^*$ is an effective Lande factor. We present the solution to the SE, Eq. (10b), corresponding to the relative coordinate motion in the next section.

### III. ORM SOLUTION TO THE RELATIVE COORDINATE SE

Separating the azimuthal coordinate part, the radial part of the SE for the relative coordinate is

$$\left\{-\frac{1}{2}\left[\frac{\partial^2}{\partial r^2}+\frac{1}{r}\frac{\partial}{\partial r}-\frac{m^2}{r^2}\right]+\frac{1}{2}\hbar^2\Omega_q^2 r^2-\hbar\Omega_q mt+\frac{1}{2}\hbar\omega_0\frac{1}{\beta}\left(1-e^{-\frac{\alpha^2\hbar^2}{4m^*}r^2}\right)\right\}\psi_{nm}(r)=\varepsilon_{n,m}\psi_{nm}(r),\qquad(15)$$

where, $\Omega_q=\frac{eB}{4m^*}$ and $\beta=\frac{\alpha^2\hbar}{m^*\omega_0}$.

A solution to the SE Eq. (15) is obtained using the ORM method. The first key step in the ORM is a transformation of the variables in the SE such that the wave function takes a Gaussian asymptotic form. Schrödinger in a paper [37] on solving eigenvalue problems by factorization pointed out the existence of such a transformation in which the Kepler problem is transformed into an oscillator problem in four dimensions. The modified SE in the new expanded space having the Gaussian asymptotic solution exhibit oscillator behavior at large distances.

In the next steps, the canonical variables (coordinate and momentum) in the transformed space are represented in terms of creation and annihilation operators $a^\dagger$ and $a$, and the Hamiltonian is written in terms of *normal ordered products* over $a^\dagger$ and $a$. The *normal order* operation (also known as Wick's transform) is an operation in which all the creation operators $a^\dagger$ are moved to the left and all the annihilation operators $a$ are moved to the right. The *normal order* operation is frequently used in quantum field theory. The Wick's transform $:q^n:$ yields the *n-th* order Hermite polynomial in $q$ which, of course, is the harmonic oscillator wavefunction apart from the exponential term. Wick's transform is in fact used in quantum field theory to eliminate infinity arising from the zero-point energy. See reference [38] for an exposition on Wick's calculus. The pure oscillator part with some yet unknown frequency, $\omega$, is extracted from the Hamiltonian written in the form $H=H_0+H_I+\varepsilon_0$, where $H_0=\omega a^\dagger a$. In addition, a requirement is imposed that the interaction part, $H_I$, does not contain terms quadratic in the canonical variables so that they are completely absorbed in the oscillator part. This requirement sets the value of the oscillator frequency.

The contribution of the interaction part of the Hamiltonian $H_I$ is obtained in the perturbation approach. It is important to note, however, that the results of the ORM using the perturbation approach for the $H_I$ is different from the standard perturbation method with the Coulomb interaction being the perturbation contribution [36]. We will show that the ORM results in the 0th order perturbation with $H_I$ are significantly better than the results obtained by the standard perturbation method. The reason for the better ORM results is that the quadratic terms arising from the interaction terms in the standard perturbation are included in $H_0$ and the solution to the SE with $H_0$ is exact. See details of the ORM solution in Appendix A.

Solving the SE Eq. (15) by the ORM we obtain the eigenvalues $\varepsilon_{n,m}$ and hence $E_{n,m}$ as

$$E_{n,m}=E_{n,m}^0+E_{n,m}^I,\qquad(16a)$$

$$E_{n,m}^0 = \hbar\omega_0 \left[ \frac{mt}{2} + (2n+|m|+1)\gamma\xi^2 + \frac{1}{\beta}\left\{1-\left(1+\frac{\beta}{2\gamma\xi^2}\right)^{-|m|-1}\right\} - \frac{|m|+1}{2\gamma\xi^2}\left(1+\frac{\beta}{2\gamma\xi^2}\right)^{-|m|-2} + \right.$$
$$\left. \frac{3}{2\sqrt{2}}k\sqrt{\gamma}\,\frac{\Gamma(|m|+\frac{1}{2})}{\Gamma(|m|+1)} \right], \tag{16b}$$

$$E_{n,m}^I = \hbar\omega_0 \left[ k\sqrt{\frac{\gamma}{2}}\,\xi\,\frac{\Gamma(n+1)\Gamma(|m|+1/2)}{\sqrt{\pi}\,\Gamma(n+|m|+1)}\sum_{p=1}^{n}\sum_{j=\max\{p,2\}}^{2p}(-1)^j 2^{2p-j}\frac{\Gamma(j+\frac{1}{2})}{\Gamma(j+|m|+1)}\frac{\Gamma(n+j+|m|+1-p)}{\Gamma(n-p+1)\Gamma(2p-j+1)\Gamma^2(j-p+1)} - \right.$$
$$\left. \frac{1}{\beta}\,\frac{\Gamma(n+1)}{\sqrt{\pi}\,\Gamma(n+|m|+1)}\sum_{p=1}^{n}\sum_{j=\max\{p,2\}}^{2p}(-1)^j 2^{2p-j}\frac{\tau^{2j}}{(1+\tau^2)^{j+|m|+1}}\frac{\Gamma(j+\frac{1}{2})}{\Gamma(j+|m|+1)}\frac{\Gamma(n+j+|m|+1-p)}{\Gamma(n-p+1)\Gamma(2p-j+1)\Gamma^2(j-p+1)} \right], \tag{16c}$$

where, $\tau^2 = \frac{\beta}{2}\sqrt{\frac{\omega_0}{2\gamma}}\frac{1}{\xi}$ and $\xi$ is determined by solving the following equation arising from the *oscillator representation condition*

$$\xi^4 + \frac{k}{\sqrt{2\gamma}}\frac{\Gamma(|m|+\frac{1}{2})}{\Gamma(|m|+2)}\xi^3 - \frac{1}{\gamma^2}\left(1+\frac{\beta}{2\gamma\xi^2}\right)^{-|m|-2} - \frac{\gamma^2-1}{\gamma^2} = 0. \tag{16d}$$

In the limit of $\alpha \to 0$ (*i.e.*, $\beta \to 0$) while $V_0\alpha^2$ is held constant at $\frac{1}{2}m^*\omega_0^2$ our model approaches the harmonic potential model. Accordingly, in this limit, the energy eigenvalue given by Eq. (16) is consistent with the results given by Eq. (18) of Dineykhan and Nazmitdinov[1] [36].

We are mostly interested in the energy levels for $n=0$ states. It is evident from Eq. (18) that the first order perturbation contributions for the $n=0$ states are 0. Therefore, the inaccuracy of the ORM results for these states is only in the second and higher order perturbation contributions.

In the next section we present and use the ORM results for a single electron in the Gaussian confining potential to determine the parameters $\alpha$ and $\hbar\omega_0$ characterizing the Gaussian potential.

## IV.  SINGLE ELCTRON CASE

Following the ORM approach we obtain eigenvalues for a single electron in a QD characterized by the Gaussian confining potential given by Eq. (1) as

$$_sE_{n,m} = {}_sE_{n,m}^0 + {}_sE_{n,m}^I + {}_sE_{s,m}, \tag{17a}$$

$$_sE_{n,m}^0 = \hbar\omega_0\left[\frac{mt}{2}+(2n+|m|+1)\gamma\xi^2+\frac{1}{\beta}\left\{1-\left(1+\frac{\beta}{2\gamma\xi^2}\right)^{-|m|-1}\right\}-\frac{|m|+1}{2\gamma\xi^2}\left(1+\frac{\beta}{2\gamma\xi^2}\right)^{-|m|-2}\right], \text{and} \tag{17b}$$

$$_sE_{s,m} = \hbar\omega_0 \cdot \frac{1}{4}g^*\frac{m^*}{m_e}t, \tag{17c}$$

And the equation for $\xi$ is given by

$$\xi^4 - \frac{1}{\gamma^2}\left(1+\frac{\beta}{2\gamma\xi^2}\right)^{-|m|-2} - \frac{\gamma^2-1}{\gamma^2} = 0. \tag{17d}$$

---

[1] It appears that there is a typographical error of a factor of ½ in the reference.

The $_sE_{n,m}^I$ contribution is identical to that in Eq. (16c) except that the first term involving the Coulomb interaction is zero.

We use Eq. (17) for the ground state, $n = m = 0$, to fit with the single-electron data obtained from Fig. 2 of the paper by Ashoori et al. [31]. The parameter values for the Gaussian potential obtained from the data fit are: $\alpha = 0.43$ and $\hbar\omega_0 = 6.32\ meV$. We also fit the same data with the single-electron energy eigenvalues for the harmonic potential [2]. The single parameter from the harmonic potential fit is $\hbar\omega_0 = 5.4\ meV$. A slightly better fit for the harmonic potential is with $\hbar\omega_0 = 5.27\ meV$. However, given the inaccuracy of the data we estimated from FIG.2 of Ashoori et al., we fixed $\hbar\omega_0$ at 5.4 $meV^2$ that Ashoori et al. determined. The standard deviations for the least square fits are 0.08 and 0.22 for the Gaussian and the harmonic potentials, respectively. The harmonic and Gaussian potentials and the corresponding energy data fits are shown in Fig.1 and Fig. 2, respectively.

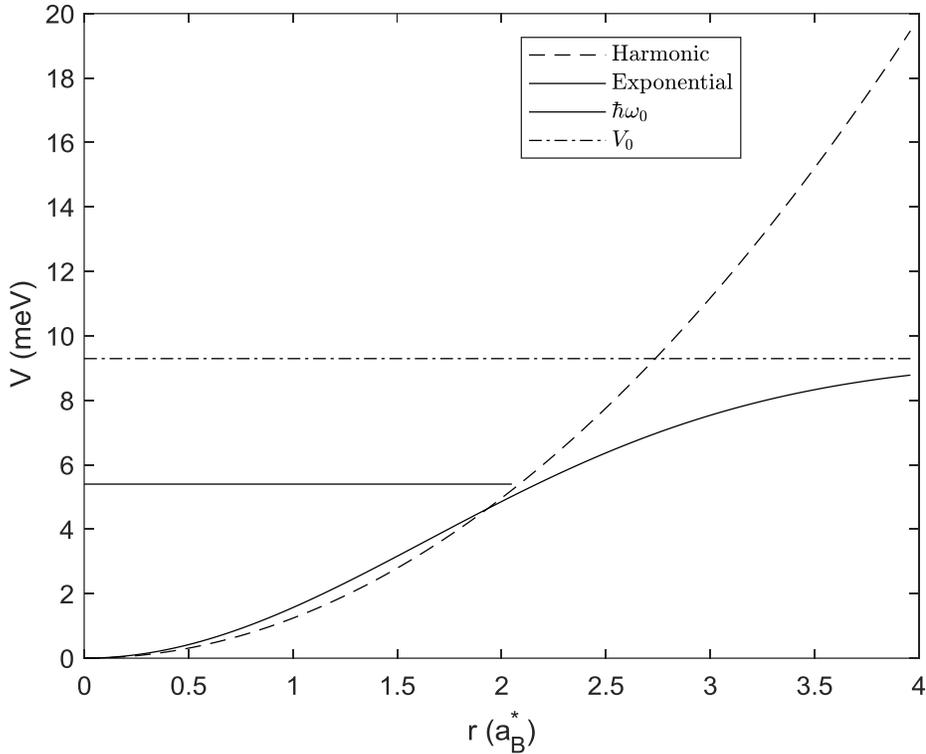

FIG. 1. Potentials in meV for the GaAs QD sample of Ashoori et al. as a function of the radius $r$ in units of effective Bohr radius. The dashed and the solid curves are the harmonic infinite potential and the finite Gaussian potential given in Eq. (2), respectively. The solid and the dash-dotted horizontal lines are the values of $\hbar\omega_0$ and $V_0$ obtained from the data fit shown in FIG. 2 for the Gaussian potential, respectively.

---

[2] We should note that the energy scale in the inset of Fig. 2 is shown as 5 meV. However, the single electron energy appears to be less than 5.4 meV if the energy origin is at 0. We assume that the single electron energy at B=0 is 5.4 meV as stated by Ashoori et al.

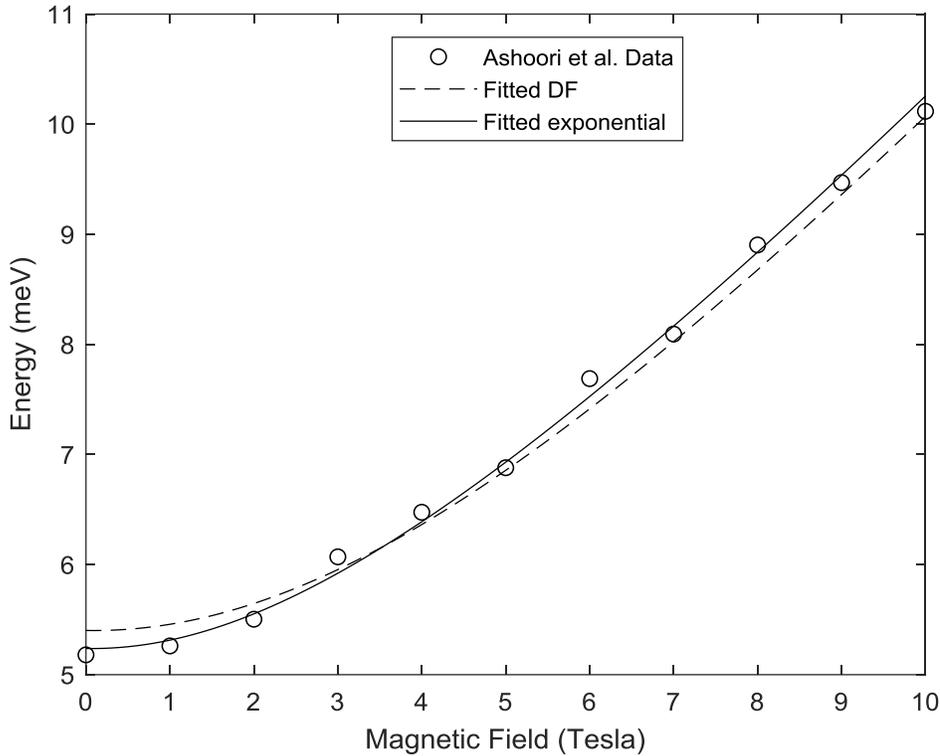

FIG. 2. Single electron energy level in meV for a 2D GaAs QD sample as a function of the magnetic field B in Tesla. The circles are the data measured by Ashoori et al. [31] using single electron capacitance spectroscopy and estimated from FIG. 2 in their paper. The dashed and the solid curves are the least square fits of the data with Darwin-Fock results for a harmonic infinite potential and the finite Gaussian potential in Eq. (2), respectively.

Now we concentrate on the two-electron energy levels in the Gaussian potential with the parameters $\hbar\omega_0$ and $\alpha$ determined above.

## V. TWO-ELECTRON RESULTS

The analytic results presented in this paper are obtained in perturbation expansion albeit better than the results obtained by standard perturbation approach. Therefore, it is important to determine the accuracy of the results and the applicability for specific QD samples.

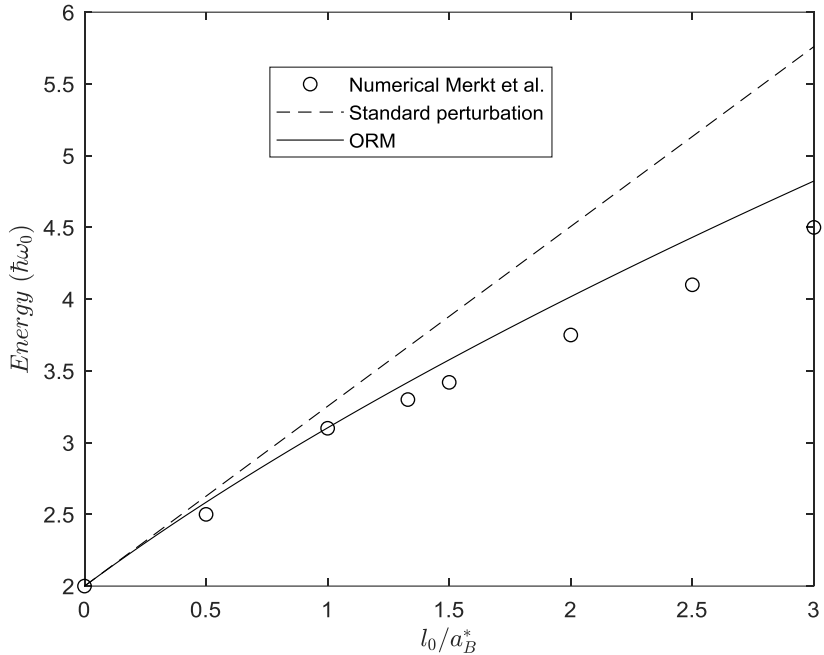

FIG. 3. The energy levels of two electrons in a harmonic confining potential in units of $\hbar\omega_0$ as a function of the oscillator length in units of the Bohr radius $a_B^*$. The circles represent the results of numerical calculations by Wagner et el. [8] (estimated from FIG. 1(b)). The dashed line represents the energy in which the Coulomb energy is obtained by standard first-order perturbation calculation. The solid line represents the results obtained by the ORM method for the harmonic potential.

Shown in FIG. 3 is a comparison of the results for a harmonic confining potential including the Coulomb interaction for the $n = m = N = M = 0$ state obtained by three methods: numerical diagonalization of the Hamiltonian by Wagner et al. [8], standard perturbation method and the ORM [36]. The results are for the magnetic field $B = 0$. As Merkt et al. [4] pointed out, indeed, the energy obtained by the standard perturbation method already starts to deviate from the numerical solution at $l_0/a_B^* = 1$. However, the ORM method provides good results at least up to $l_0/a_B^* = 2$ where the error is about 6%. For states with higher values of $|m|$ the ORM is significantly better. For $m = \pm 1$ the energy values by the standard perturbation method and the ORM are 4.70 and 4.88, respectively, compared to 4.7 (estimated from Merkt et al. FIG. 1(b)) in units of $\hbar\omega_0$ obtained by the numerical method. For a higher magnetic field $\omega_c/\omega_0 = 5$ the energy levels are 10.35 and 7.8 $\hbar\omega_0$ for the $m = 0, \pm 1$ states, respectively, obtained by the ORM method compared to the corresponding values 10.1 and 8 $\hbar\omega_0$, respectively, obtained by Wagner et al.

Using for GaAs $= l_0/a_B^* = 3.272/\sqrt{\hbar\omega_0}$ , we obtain $k = 1.3$. With this value of $k$ we calculate the energy levels as a function of the magnetic field from Equations (11), (13), (14), and (16). At this value of $k$ we expect the ORM results to be good. The energy levels for $m = 0, -1, -2 \ and -3$ states are shown in FIG. 4. The potentials are characterized by the parameters $\alpha = 0.43 \ and \ \hbar\omega_0 = 6.32 \ meV$ as discussed earlier. For comparison with infinite harmonic potential the corresponding results with $\beta \rightarrow 0$ are shown as the dotted curves. The singlet to triplet crossing for the ground state is at 2.3 T compared to 3.1 T for the infinite harmonic potential. Ashoori et al. [31] determined, from the "bump" of the two-electron ground state energy, the magnetic field value for the singlet-to-triplet transition to be at 1.5 T.

According to the calculation by Wagner et al. the crossing is at 3.6 T. Ashoori et al. conjectured that the discrepancy may be due to the assumption of strictly parabolic potential for the QD. Even though we expect the finite Gaussian potential to represent the QD better than the harmonic potential it is unlikely that even an exact solution would match perfectly with the experimental data for the singlet-to-triplet crossing. However, we believe that even though our model involves approximations, namely approximate separation of the potential CM and relative coordinates, and the perturbation approximation involved in obtaining the analytic solution, our results validate the conjecture by Ashoori et al.

In Fig. 4 we have also plotted the energy levels as dashed lines for the Gaussian potential with an additional correction term. We had neglected $\alpha^2 R^2$ including $\alpha^4 R^2 r^2$ to obtain the confining potential separable in CM and RC. With the solution of the RC SE we can now obtain the expectation value of the $\alpha^4 R^2 r^2$ term for the $|n, m>$ state and use it as a correction term to the CM potential. The correction term, of course, is dependent on the $n, m$ quantum numbers of the RC state. The effective potential given by Eq. (5) is modified as

$$V = 2V_0[1 - e^{-\alpha^2 r^2/4} + \alpha^2 R^2 < n, m | e^{-\alpha^2 r^2/4} | n, m >]. \tag{18}$$

In particular, for the $n = 0$ states for which the wave function is $\sqrt{\frac{\omega_0 \gamma}{2\pi}} \xi e^{-\frac{\omega_0 \gamma}{4} \xi^2 q^2}$, the potential is written as

$$V = 2V_0[1 - e^{-\alpha^2 r^2/4} + \zeta_m \alpha^2 R^2], \tag{19}$$

where, $\zeta_m = \left(1 + \frac{\beta}{2\gamma \xi^2}\right)^{-(|m|+1)/2}$.

Note that $\zeta_m$ is dependent on $m$ via Eq. (16d) for $\xi$. For this potential the $m$-dependent CM energy is given by

$$E_{N,M}(m) = 2\hbar \omega_0 \zeta_m \left[\gamma(2N + |M| + 1) + \frac{1}{2} Mt\right]. \tag{20}$$

The dashed curves in Fig. 4 are the energy values in which the $E_{N,M}$ is given by Eq. (20). The energy components, $E_{n,m}$ and $E_s$, of course, are not affected by the CM potential correction.

With the CM potential correction, the singlet-to-triplet ground state energy transition occurs at 1.9 T compared to the experimental value at 1.5 T. The closer agreement with the experimental value further highlights the importance of the finite potential compared to infinite harmonic potential for 2D QDs.

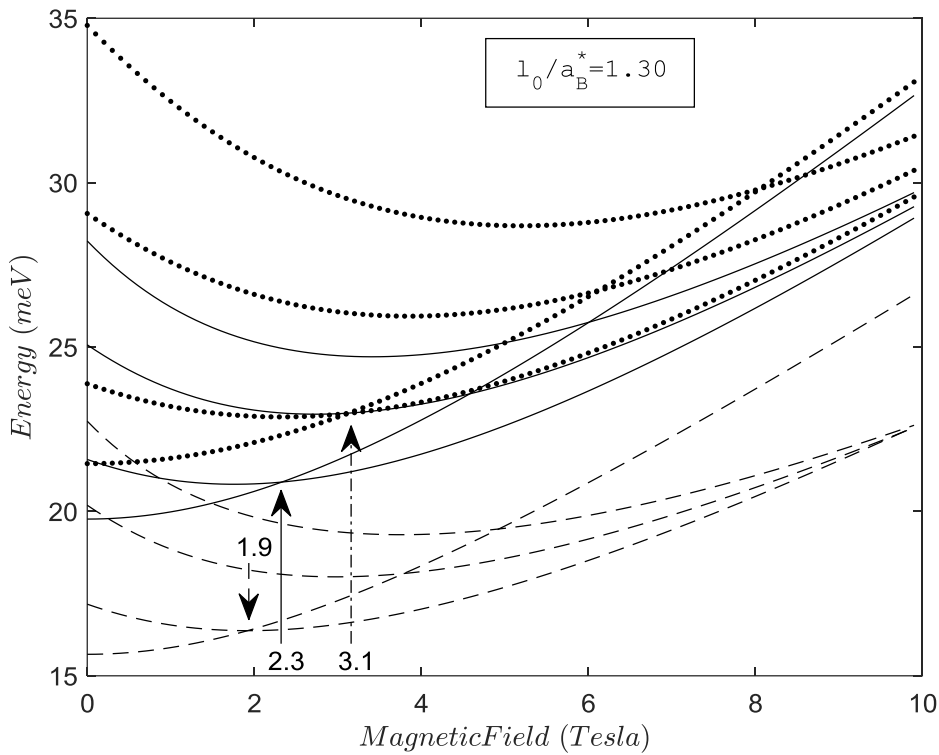

FIG. 4. The energy levels for $m = 0, -1, -2, -3 \ and - 4$ states as a function of the magnetic field. The solid lines represent the results for the Gaussian potential given by Eq. (5), the dashed lines represent the results for the corrected Gaussian potential given by Eq. (19), and the dotted lines represent the results for the harmonic potential. The energy level curves at zero magnetic field are in increasing order for the $m = 0, -1, -2, -3 \ and - 4$ states. The singlet-to-triplet transitions are shown by the arrows at 1.9 T, 2.3 T and 3.1 T for the corrected Gaussian, Gaussian, and the harmonic potentials, respectively.

In FIG. 5 we have plotted the calculated and experimental ground state energy versus the magnetic field. The solid line represents the results for the Gaussian potential calculated by using Equations 16(a-d). We should note that the slope of the function of $\xi$ in Eq. (16d) is small near the zero crossing. As a result, it is difficult to obtain accurate solution to Eq. (16d), particularly for large $|m|$ and small $B$ values. The dashed line represents the results for the corrected Gaussian potential given by Eq. (19) and normalized with the $B = 0$ energy value for the corresponding uncorrected Gaussian value. The circles represent the experimental data estimated from FIG. 2 of reference [31] and normalized with the calculated value at $B = 0$. While the theoretical ground state energy agrees well with experimental data, they diverge continuously as the magnetic field increases. This discrepancy with the theoretical results cannot be attributed to the approximations involved in the calculation. As discussed earlier, the agreement of our results with the results of the numerical calculation by Wagner et al. [8] is better at higher magnetic fields. The divergence would be even more if we use the harmonic potential.

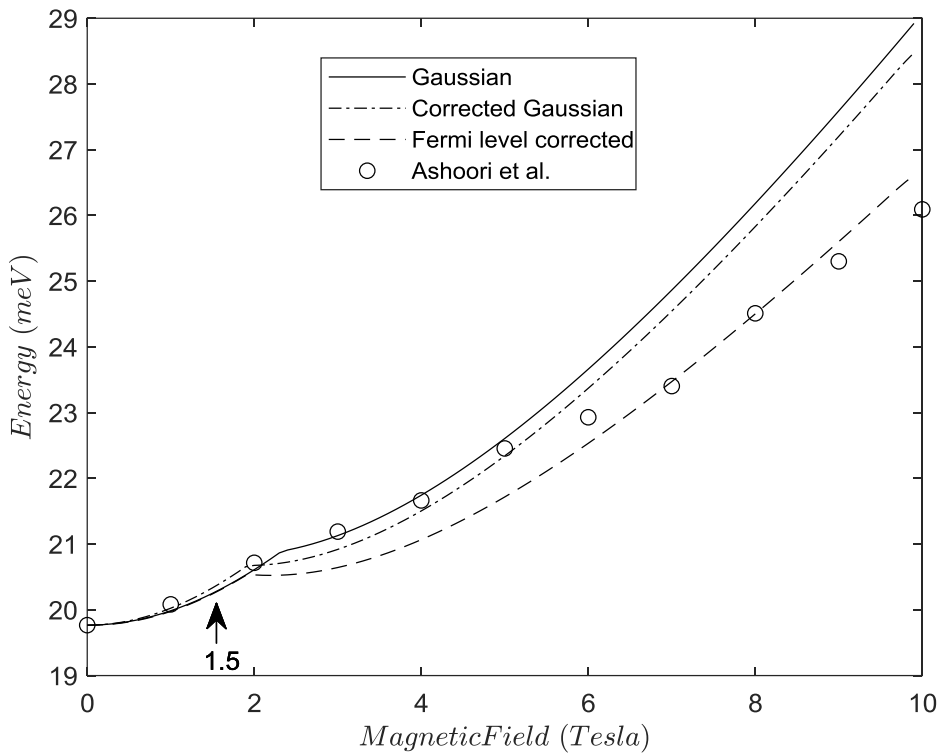

FIG. 5. Ground state energy level as a function of the magnetic field. The solid line represents the results calculated for the Gaussian potential. The dash-dotted line represents the results calculated for the corrected Gaussian potential. The dashed line represents the results with the electrode Fermi level correction. The circles are the estimated experimental data from FIG. 2. of Ashoori et al. [31]. The arrow indicates the singlet-to-triplet crossing according to the experiment.

As pointed out by Ashoori [39], one potential factor contributing to the divergence at higher magnetic fields is the shift in the Fermi energy of the n$^+$ electrode with the magnetic field. The energy levels are measured through the resonance of the QD energy levels with the electrode Fermi energy. The Fermi energy itself increases as the magnetic field increases. At as high a magnetic field as 10 T and low temperature ($T = 0.35\,K$) one would expect the Fermi level to undergo de Haas-van Alphen (dHvA) oscillation. However, there is no discernible oscillation in the experimental data. We believe that the oscillation is dampened by the broadening of the Landau levels due to doping ion and electron-electron interactions. Eisenstein et al. [40] and Ashoori and Silsbee [41] obtained the density of states of 2D electrons in a GaAs-GaAlAs heterostructure at high magnetic fields and have shown that the Landau levels are significantly broadened. Eisenstein et al. used a theoretical model with Gaussian distribution and Ashoori et al. used a Lorentzian distribution of the DoS around the Landau levels to fit their experimental data. We have not succeeded in obtaining a closed form solution for the Fermi energy as a function of the magnetic field with Landau level broadening. Instead we have used the following ansatz for the Fermi energy as a function of the magnetic field

$$\mu = \mu_0 \left[ 1 + \frac{3}{16} \left( \frac{\hbar \omega_c}{\mu} \right)^2 \left( \frac{m^{*2} g^2}{8 m_e^2} - \frac{1}{6} \right) - \delta \left( \frac{\hbar \omega_c}{\mu} \right)^{3/2} \right]^{-2/3} \tag{21}$$

where, $\mu$ is the B-dependent Fermi energy, $\mu_0 = \frac{\hbar^2}{2m^*}(3\pi^2 n)^{2/3}$ is the Fermi energy at $B = 0$, $n$ is the electron density, and $\delta$ is a parameter. The Fermi energy $\mu_0$ of the n$^+$ electrode with a doping concentration $4 \times 10^{17}\, cm^{-3}$ [31] is 11.74 meV. In this ansatz we have replaced the oscillation factor in the $\left(\frac{\hbar\omega_c}{\mu}\right)^{3/2}$ dependent dHvA oscillation term [41] with the constant parameter $\delta$. See Appendix B for a justification for the ansatz. The dashed curve in Fig.5 represents the calculated 2-electron g.s. energy with the Fermi energy correction given by Eq. (21). The value of the parameter $\delta$ is 0.075. In comparison, the value of the coefficient of the first term in the sum of the infinite series for the dHvA oscillation without any Landau level broadening is 0.169.

Additional factors may contribute to the divergence of the g.s. energy level at higher magnetic fields. For example, the effect of screening from the tunneling electrode may be important [39]. Bruce and Maksym [30] found that the effect of screening for the specific geometry of the device they used for their calculation is insignificant. They noted, however, that the effect could be important for different device geometry. Given that the electrode separation in the experimental device of Ashoori et al. [31] is larger the screening effect may not be significant for this device. Another possibility is the next order perturbation term we have neglected in the ORM interaction. As mentioned earlier, the first order interaction perturbation term given by Eq. (16c) is zero for $n = 0$. However, the next order term, though small, involves a factor $\gamma^2$ and that may be significant at higher magnetic field to account for a non-negligible contribution to the divergence. Further investigation is required to determine whether these contributions are significant.

## VI.  SUMMARY

Analytic expressions for the energy levels of a 2-electron system in a 2D QD modeled with a finite Gaussian potential and subjected to a magnetic field is obtained. In the appropriate limits the results are shown to match very well with previous analytical and numerical results. Using the expressions, the magnetic field at which the singlet to triplet ground state crossing occurs is calculated. The calculated value is closer to the experimental value compared with the infinite harmonic potential results in previous theoretical models. It may be possible to use the ORM wave functions to calculate and diagonalize the Hamiltonian matrix to obtain more accurate numerical results for the finite Gaussian potential. The ground state energy is also calculated as a function of the magnetic field. The divergence of the theoretical value increases continuously with the magnetic field. The g.s. energy level matches very well with the theoretical results when a correction for the electrode Fermi energy dependence on the magnetic field is incorporated.

### ACKNOWLEDGMENTS

It is a pleasure to thank R. Ashoori for his encouragement and several insightful comments, in particular, for suggesting the electrode Fermi level shift as a contributing factor in the divergence of the calculated g.s. energy levels from the experimental values at higher magnetic fields.

**APPENDIX A: ORM - Oscillator Representation Method**

The requirement that the pure oscillator part of the Hamiltonian is absorbed in the harmonic oscillator Hamiltonian $H_0$ leads to the to the condition

$$\frac{\partial \varepsilon_0}{\partial \omega} = 0, \tag{A1}$$

which determines $\omega$, the oscillator frequency. This condition is known as *the oscillator requirement condition* (ORC).

Dineykhan et al. [32] used the following transformations that provide the Gaussian asymptotic wave function behavior[3]:

$$r = q^{2\rho} \tag{A2}$$

$$r\psi(r) = q^a \phi(q). \tag{A3}$$

The radial part of the SE in the new variable $q$ in the expanded space is then obtained as the following:

$$\left[ -\frac{1}{2}\left( \frac{\partial^2}{\partial q^2} + \frac{d-1}{q}\frac{\partial}{\partial q} \right) + W_l(q; E) \right] \phi(q) = \varepsilon(E)\phi(q), \tag{A4}$$

where

$$W_l(q; E) = -\frac{K(l,\rho,d)}{2q^2} + 4\rho^2 (q^2)^{(2\rho-1)}(V(q^2) - E), \tag{A5}$$

$$d = 2a - 2\rho + 2, \tag{A6}$$

$$K(l,d) = \frac{1}{4}[(d-2)^2 - 4(2l+1)^2]. \tag{A6}$$

It should be noted here that $d$ is the dimension of the hyperspace, the energy $E$ is now incorporated in the new potential $W_l$, and $q^2 = q_i q_i$ summed over the repeated index $i$ from 1 to the dimension $d$. We also note that the SE is written in dimensionless form by using the length unit as $\hbar^2/me^2$ and the energy unit as $me^4/\hbar^2$. An interesting aspect of the ORM is that the dimension of the hyperspace can be a variation parameter and as a result can be non-integer. Even though an integer dimension is used to derive the energy equations, the dimension appears in the end results just as a parameter and thus can be varied to obtain the energy minimum. However, in our calculation we have set $\rho = 1$.

---

[3] We should note that in the context of the ORM $\rho$ is a parameter while $\rho$ was used as the scaled radius variable $r$ in the analytic model section.

The energy spectrum $E_{nl}$ of the original system is obtained from the radial excitation spectrum $\varepsilon^{[n_r]}$ of the Hamiltonian of Eq. (11)

$$H(E_{nl})\phi^{[n_r]}(q) = \varepsilon^{[n_r]}\phi^{[n_r]}(q), \quad (n_r = 0,1,2,\dots), \tag{A7}$$

and it is determined by

$$\varepsilon^{[n_r]}(E_{nl}) = \varepsilon^{[n_r]}(l,d;E) = 0, \tag{A8}$$

where $n_r$ is the radial quantum number.

If the potential $V(r)$ does not have a repulsive character as $r_0 \to 0$, $K(l,\rho,d)$ is typically chosen to be zero and that leads to the equation for the dimension,

$$d = 4l + 4. \tag{A9}$$

The oscillator representation is then obtained by writing the Hamiltonian $H$ in the form

$$H = \tfrac{1}{2}(p^2 + \omega^2 q^2) + \left(W_l(q) - \tfrac{1}{2}\omega^2 q^2\right), \tag{A10}$$

and by introducing the usual creation and annihilation operators, $a^\dagger$ and $a$, respectively, in terms of the canonical variables $q_j$ and $p_j$ $(j = 1,2,\dots,d)$, as

$$q_j = \tfrac{1}{\sqrt{2\omega}}\left(a_j + a_j^\dagger\right), \quad p_j = \tfrac{1}{i}\sqrt{\tfrac{\omega}{2}}\left(a_j - a_j^\dagger\right). \tag{A11}$$

The creation and annihilation operators satisfy the standard commutation relation

$$\left[a_j, a_k^\dagger\right] = \delta_{jk}, \quad j,k = 1,2,\dots d. \tag{A12}$$

After *normal ordering the products* over $a^\dagger$ and $a$, in which all the creation operators are moved to the left and the annihilation operators are moved to the right as if they commute and some manipulation the ORM Hamiltonian is obtained as

$$H = H_0 + H_I + \varepsilon_0, \tag{A13}$$

where

$$H_0 = \omega a_j^\dagger a_j, \tag{A14}$$

$$H_I = \int_{-\infty}^{\infty}\left(\tfrac{dk}{2\pi}\right)^d \widetilde{W}(k)\exp\left(\tfrac{-k^2}{4\omega}\right) : e_2^{i(kq)} :, \quad \text{and} \tag{A15}$$

$$\varepsilon_0 = \tfrac{d\omega}{4} + \int_{-\infty}^{\infty}\left(\tfrac{dk}{2\pi}\right)^d \widetilde{W}(k)\exp\left(-\tfrac{k^2}{4\omega}\right). \tag{A16}$$

As mentioned earlier, the symbol :*: represents the *normal ordering* of the products over $a^\dagger$ and $a$, and in $kq = k_j q_j$ as well as in $k^2$ and $q^2$ summation over repeated indices are assumed. The function $e_2^{i(kq)} = e^{i(kq)} - 1 + \frac{k^2 q^2}{2d}$. The function $\widetilde{W}$ is the Fourier transform of $W(q)$ in $d$-dimension and given as

$$\widetilde{W}(k) = \int_{-\infty}^{\infty} (d\xi)^d W_l(\xi) \exp(ik\xi), \quad k\xi = k_j \xi_j. \tag{A17}$$

The energy spectrum is then obtained by calculating the contribution of the interaction part $H_I$ of the Hamiltonian in the perturbation approach. In the zeroth order approximation the energy spectrum is determined by

$$2n_r\omega + \langle n_r | H_I | n_r \rangle + \varepsilon_0 = 0. \tag{A18}$$

The equation for $\varepsilon_{n,m}^0$ obtained from Eq. (A16) is

$$\varepsilon_0 = \hbar\omega \frac{d}{4} \left(1 + \frac{\Omega_q^2}{\omega^2}\right) - \left(\varepsilon_{n,m}^0 - \hbar\omega \frac{mt}{4}\right) + \frac{\hbar\omega_0}{2\beta}\left[1 - \left(1 + \frac{\beta}{4}\frac{\omega_0}{\omega}\right)^{-\frac{d}{2}}\right] + \hbar\omega_0 \frac{k}{2}\left(\frac{\omega}{\omega_0}\right)^{1/2} \frac{\Gamma(\frac{d-1}{2})}{\Gamma(\frac{d}{2})}. \tag{A19}$$

The interaction part of the Hamiltonian is obtained from Eq. (A15) as

$$H_I = \hbar\omega_0 \frac{k}{2}\left(\frac{\omega}{\omega_0}\right)^{1/2} \int_{-\infty}^{\infty} \frac{d\tau}{\pi} \int_{-\infty}^{\infty} \left(\frac{d\eta}{\sqrt{\pi}}\right)^d e^{-\eta^2(1+\tau^2)} : e_2^{-2i\sqrt{\hbar\omega}\tau\boldsymbol{\eta}.\boldsymbol{q}} :$$

$$- \hbar\omega_0 \frac{\beta}{2}\int_{-\infty}^{\infty} \left(\frac{d\eta}{\sqrt{\pi}}\right)^d e^{-\eta^2\left(1+\frac{\beta\omega_0}{4\omega}\right)} : e_2^{-i\sqrt{\hbar\omega_0}\beta\boldsymbol{\eta}.\boldsymbol{q}} : . \tag{A20}$$

We obtain Eq. (16b) from Equations (A1) and (A19). Equations (16c) and (16d) are obtained from Eq. (A20) and (A18), respectively.

See reference [32] for the details of the calculation of the radial eigenstates $|n_r\rangle$ and the matrix elements $\langle n_r | H_I | n_r \rangle$. We obtain Eq. (16d) for the oscillator frequency $\omega$ from Eq. (A1) for the Gaussian potential in the RC Hamiltonian given in Eq. (8). The interaction energy given in Eq. (16c) is obtained by using Eq. (A18) for the Gaussian potential with the frequency obtained by solving Eq. (16d). The eigen functions $|n_r\rangle$ that include the oscillator part of the Coulomb interaction may provide a good basis set for numerical calculations of the energy levels with the Gaussian potential.

As mentioned earlier the divergence of the calculated energy values with the experimental values at higher magnetic fields may be accounted for by the higher order terms in $\langle n_r | H_I | n_r \rangle$. Having the wave functions and energy levels calculated in the zeroth order the next order term in the eigen energies can be calculated using the equation:

$$E_n^{(2)} = -\left\langle \psi_n^{(0)} \middle| [H_I - E_n^1] \frac{1}{H_0 - 2n\omega} \middle| [H_I - E_n^1] \right\rangle. \tag{A19}$$

## APPENDIX B: Magnetic Field Dependent Fermi Energy

With no Landau level broadening the Fermi energy has an oscillatory term as shown in the following equation [41]

$$\mu = \mu_0 \left[ 1 + \frac{3}{16}\left(\frac{\hbar\omega_c}{\mu}\right)^2 \left(\frac{m^{*2}g^2}{8m_e^2} - \frac{1}{6}\right) - \frac{3\pi KT}{\hbar\omega_c}\left(\frac{\hbar\omega_c}{\mu}\right)^{3/2} \sum_{n=1}^{\infty} \frac{(-1)^n}{\sqrt{n}} \frac{\sin\left(\frac{2\pi n\mu}{\hbar\omega_c} - \frac{\pi}{4}\right)\cos(n\pi gm^*/2m_0)}{\sinh(2\pi^2 nKT/\hbar\omega_c)} \right]^{-2/3} \quad (B1)$$

This formula is derived by setting $\frac{\partial F}{\partial \mu} = 0$ and the free energy $F$ is calculated by using the equation,

$$F = N\mu + \int_0^{\infty} z(E)\frac{\partial f}{\partial E}\,dE, \tag{B2}$$

where,

$$z(E) = \frac{1}{2\pi i}\int_{c-i\infty}^{c+i\infty} e^{Es} s^{-2} Z(s)\,ds, \tag{B3}$$

$$Z(\beta) = \sum_l e^{-\beta\varepsilon_l}, \tag{B4}$$

$$\varepsilon_l = \frac{\hbar^2 k_z^2}{2m^*} + \left(l + \frac{1}{2}\right)\hbar\omega_c \pm \left(\frac{g\hbar\omega_0}{4}\right), \text{ and} \tag{B5}$$

$$f = \{exp[(E - \mu)/KT] + 1\}^{-1}. \tag{B6}$$

As the summation in Eq. (B4) is carried out with the eigen energies given by Eq. (B5), a csch $\left(\frac{\beta\hbar\omega_c}{2}\right)$ factor appears in $Z(\beta)$. To carry out the integral in Eq. (B3) the integration path is replaced with a contour integral. The csch function has an infinite number of poles along the imaginary axis of the contour that leads to the oscillatory term in Eq. (B1) giving rise to dHvA effect. However, with Landau level broadening, Eq. (B4) needs to be replaced by

$$Z(\beta) = \frac{1}{\sqrt{2\pi}\gamma}\sum_l \int_0^{\infty} e^{-\frac{(\varepsilon-\varepsilon_l)^2}{2\gamma^2}} e^{-\beta\varepsilon_l}\,d\varepsilon. \tag{B7}$$

A corresponding equation for Lorentzian broadening could also be used. With this replacement, the summation in Eq. (B4) cannot be carried out to obtain the csch term. Therefore, the oscillatory term disappears from the integration in Eq. (B3). The integration along the negative real axis of the contour then adds a term arising from the broadening given by Eq. (B6). However, we could not obtain a closed form expression for the term. Instead, noting that at low temperature the $\frac{KT}{\hbar\omega_c}$ term cancels out in the oscillatory term in Eq. (B1), we assume that the corresponding term arising from the Landau level broadening is proportional to $\left(\frac{\hbar\omega_c}{\mu}\right)^{3/2}$ which we have used in the ansatz given by Eq. (21).

---

[1]    V. Fock, *Z. Phys.,* vol. 47, p. 446, 1928.

[2]    C. G. Darwin, *Proc. Cambridge Philos. Soc.,* vol. 27, p. 86, 1931.

[3]     L. Jacak, P. Hawrylak and A. Wojs, Quantum Dots, New York: Springer, 1998.

[4]     U. Merkt, J. Huser and M. Wagner, *Phys. Rev. B,* vol. 43, no. 9, p. 7320, 1991.

[5]     Q. P. Li, K. Karrai, S. K. Yip, S. Das Sarma and H. D. Drew, *Phys. Rev. B,* vol. 43, p. 5151, 1991.

[6]     T. Chakraborty, V. Halonen and P. Pietilainen, *Phys. Rev. B,* vol. 43, no. 14, p. 289, 1991.

[7]     D. Gerhardts and R. Pfannkuche, *Phys. Rev. B,* vol. 44, no. 13, p. 132, 1991.

[8]     M. Wagner, U. Merkt and A. V. Chaplik, *Phys. Rev. B,* vol. 45, p. 1951, 1992.

[9]     A. Kumar, S. Laux and a. F. Stern, *Phys. Rev. B,* vol. 42, p. 5166, 1990.

[10]    D. Broido, K. Kempa and P. Bakshi, *Phys. Rev. B,* vol. 42, p. 5166, 1990.

[11]    V. Gudmundsson and R. Gerhardts, *Phys. Rev. B,* vol. 43, p. 12098, 1991.

[12]    D. Pfannkuche, V. Gudmundsson and P. Maksym, *Phys. Rev. B.,* vol. 47, no. 4, 1993.

[13]    G. W. Bryant, *Phys. Rev. Lett.,* vol. 59, p. 1140, 1987.

[14]    P. A. Maksym and T. Chakrabarty, *Phys. Rev. Lett.,* vol. 65, p. 108, 1990.

[15]    M. Taut, *J. Phys. A.,* vol. 27, p. 1045, 1994.

[16]    B. Kandemir, *Phys. Rev. B,* vol. 72, p. 165350, 2005.

[17]    F. Prudente, L. Costa and J. Viana, *J. Chem. Phys.,* vol. 123, p. 224701, 2005.

[18]    D. Nader, J. Alvarez-Jimenez and H. Mejia-Diaz, *Few-Body Syst.,* vol. 58, p. 116, 2017.

[19]    A. Shaer, M. Elsaid and E. Hjaz, *Nanosystems: Physics, Chemsitry, Mathematics,* vol. 10, no. 5, p. 530, 2019.

[20]    D. Miserev and O. Sushkov, *Phys. Rev. B,* vol. 100, no. 20, p. 5129, 2019.

[21]    F. Caruso, V. Oguri and F. Silvera, *Physica E,* vol. 105, p. 182, 2019.

[22]    N. Johnson and M. Payne, *Phys. Rev. Lett.,* vol. 67, no. 9, p. 1157, 1991.

[23]    J. Kainz, S. Mikahailov, A. Wensauer and U. Rossler, *Phys. Rev. B,* vol. 65, p. 115305, 2002.

[24]    P. A. Maksym and N. A. Bruce, *Physica E,* vol. 1, p. 211, 1997.

[25]    O. Makarovsky, O. Thomas, A. Balanov, L. Eaves, A. Patane, R. Campion, C. Foxon, E. Vdovin, D. Maude, G. Kiesslich and R. Airey, *Phys. Rev. Lett.,* vol. 101, p. 226807, 2008.

[26]    F. Peeters and V. Schweigert, *Phys. Rev. B,* vol. 53, p. 1468, 19916.

[27]    A. Poszawa, *Few-Body Syst,* vol. 57, p. 1127, 2016.

[28]    J.-L. Zhu, Z.-Q. Li, J.-Z. Yu, K. Ohno and Y. Kawazoe, *Phys. Rev.,* vol. 55, no. 23, p. 15819, 1997.

[29]    J. Adamowski, M. Sobkowicz, B. Szfran and S. Bednarek, *Phys. Rev. B,* vol. 62, no. 7, p. 4234, 2000.

[30]    N. A. Bruce and P. A. Maksym, *Phys. Rev. B,* vol. 61, p. 4718, 2000.

[31]    R. C. Ashoori, H. L. Stormer, J. S. Weiner, L. N. Pfeiffer, K. W. Baldwin and K. W. West, *Phys. Rev. Lett.,* vol. 71, p. 613, 1993.

[32]    M. Dineykhan and G. V. Efimov, *Rep. Math. Phys.,* vol. 36, no. 287, 1995.

[33]    M. Dineykhan and G. V. Efimov, *Few-Body Systems,* vol. 16, p. 59, 1994.

[34]    M. E. Amin and M. A. El-Asser, *Brazilian Journal of Physics,* vol. 39, no. 2, p. 301.

[35]    S. Chaudhuri, *To be published.*

[36]    M. Dineykhan and R. G. Nazmitdinov, *Phys. Rev. B,* vol. 55, no. 20, p. 13707, 1997.

[37]    E. Schrödinger, *Proc. R. Irish Acad.,* vol. 46, p. 183, 1941.

[38]    A. Wurm and M. Berg, *Am. J. Phys.,* vol. 76, no. 1, p. 65, 2008.

[39]    R. Ashoori, *Private communication.*

[40]    J. Eisenstein, H. Stormer, V. Narayanamurti, A. Cho, A. Gossard and C. Tu, *Phys. Rev. Lett.,* vol. 55, no. 8, p. 875, 1985.

[41]    R. C. Ashoori and R. H. Silsbee, *Solid State Commun.,* vol. 81, no. 10, p. 821, 1992.

[42]    J. Callaway, Quantum Theory of Solid State, New York: Academic Press, Inc., 1976.